\documentclass[aps,prl,nofootinbib]{revtex4}

\usepackage[utf8x]{inputenc}
\usepackage{amsmath}
\usepackage{amssymb}
\usepackage{amsfonts}
\usepackage{graphics}
\usepackage[pdftex]{graphicx}
\usepackage[normal]{subfigure}
\usepackage{color}

\begin{document}

\title{On parameter identifiability problem in Agent Based economical models}
\author{Giuseppe Di Molfetta$^*$}
\affiliation{Master 2 Recherche Economie Theorique et Empirique, Universite Paris 1 Pantheon - Sorbonne}
\affiliation{Supervised by Jean-Bernard Chatelain (Director) and Antoine Mandel (Co-Director)}
\date{\today}

\vspace{1cm}

\begin{abstract}

Identifiability of parameters is a fundamental prerequisite for model identification. It concerns uniqueness of the model parameters determined from experimental or simulated observations. This dissertation specifically deals with structural or a priori identifiability: whether or not parameters can be identified from a given model structure and experimental measurements. We briefly present the identifiability  problem in linear and non linear dynamical model. We compare DSGE and Agent Based model (ABM) in terms of identifiability of the structural parameters and we finally discuss limits and perspective of numerical protocols to test global identifiability in case of ergodic and markovian economical systems. 
\end{abstract}

\maketitle

\vspace{1cm}

\section{Introduction}
\begin{large}

\textit{When the crisis came, the serious limitations of existing economic and financial models immediately became apparent [...] Macro models failed to predict the crisis and seemed incapable of explaining what was happening to the economy in a convincing manner. As a policy-maker during the crisis, I found the available models of limited help. In fact, I would go further: in the face of the crisis, we felt abandoned by conventional tools. \\
\vspace{0.25cm}\\
We need to deal better with heterogeneity across agents and the interaction among those heterogeneous agents. We need to entertain alternative motivations for economic choices. [...] Agent -based modeling dispenses with the optimization assumption and allows for more complex interactions between agents. Such approaches are worthy of our attention. \\}
\vspace{0.1cm}\\
Jean-Claude Trichet,  2010
\vspace{1cm}

Agent Based Models (ABM) are generally introduced in the context of complex social phenomena with the aim to modeling the emergence of complexity from dynamic interaction rule based agents (see for exemple \citet{axelrod1997complexity} \citet{grimm2006standard}). Recently ABMs have been introduced in macroeconomics (\citet{colander2008beyond, dosi2013income, gualdi2015tipping, gualdi2015monetary})  in contrast with the Representative Agent (RA) models such as the Dynamic Stochastic General Equilibrium models (DSGE).\\

We know that in DSGE the economic agents are assumed to be identical, non interacting, rational agents, the models are simple enough to lead to closed form analytical results, with simple narratives and well-trodden calibration avenues. We can't say the same for ABMs. In fact in the latter case the high dimensional parameters spaces and explicit or implicit choices of behavioral rules is so large that the results of the model may appear unreliable and arbitrary.  \\

In ABMs the economic system is composed by many different autonomous agents that interact with each other and with the environment. The result is a system that exhibits emergent properties: the properties at the macro level cannot be explained directly by the properties at the micro level. \\
Agent based modeling is a tool used to overcome the limitations of pure mathematical analysis, it allows the construction of more realistic models. The main interest of ABMs is that they are very versatiles and can provide extremely valuable tools for generating scenarios, that can be used to test the effect of policy decisions in times of large economic instabilities.\\

But an ABM suffers from the lack of a well defined set of equations because from a numerical model is hard to infer a representation as a system of dynamical equations. Indeed the fundamental difference between DSGE and ABM is that in the former we can define explicitly a functional, that is a relationship between a dependent variable and explanatory variables.
In ABM this relationship is implicitly defined in the numerical code. \\ In addition to this in ABM a single run does not provide any information on the robustness of the solutions. One way to treat such a problem in agent computing is through multiple runs, systematically varying initial conditions or parameters in order to asses the robustness of results (\citet{axtell2000agents}). 


This allows ABMs' researchers to the hard problem of standard estimation methods in order to offers a true real and feasible alternative to currents economical models. In the following we focus on the identifiability  problem, well known for a very large class of dynamical systems, DSGE included and never studied systematically for Agent Based model. \\

The dissertation is organized as follows: in the first section we briefly overview the problem of structural or a priori parameter identifiability in linear and non linear models; in the second section we provide the analogy between DSGE and AB model that allows us to compare problems and solution in both cases; in the last section we propose two simple protocols to test identifiability in a very large class of economical models, the ergodic and markovian systems and finally discuss a mean field approach.

\vspace{1cm}

\section{The problem of the identifiability}


In general the problem of determining the parameter values of a system from input-output data is called the estimation problem. The identifiability  problem is more circumscribed: given a model of the system and specific input-output experiments we ask if the parameters of the model can be uniquely determined. \\

The identifiability problem is not specific to econometrics but spreads from statistics to control and system engineering, from chemical to biological problems. Theoretically it is well understood at least in linear dynamical model. First studies appears with (\citet{koopmans1950identification}) and (\citet{pesaran1981identification}) with a specific analysis on rational expectation economic model. The simplest but not trivial model on which identifiability  was investigated rigorously, was the simultaneous equations problem (see e.g. \citet{pindyck1998econometric}). \\


According to Canova (\citet{canova2009back}) there are four classes of identification problem that may arise: (i) The absence of a unique minimum in objective function 
\footnote{In general to estimate the parameters is sufficient to choose the value of the parameters that minimize the distance between the theoretical (simulated ) moments and the observed moments. The general expression of the objective function to be minimized can be found in Gourieroux and Monfort, where also the asymptotic properties of the estimator are shown. The objective function transcribe in general in: $J(\delta,W) = (\mu^R - \mu^T(\delta))' W (\mu^R - \mu^T(\delta))$
where $\mu^R$ is the vector of dimension M containing the first M non-centered moments computed over the real data, $\mu^T$ is the vector of dimension M containing the first M non-centered moments computed over theoretical (simulated) data. In the case of simulated data, the moments depends on the parameters used to run the simulation. The element $m$ of the vector $\mu^T$ is $ \mu^T_m = \frac{1}{S} \sum_{s=1}^S(\frac{1}{n} \sum_{t=1}^n y_t^m)_s$.  The estimated set of parameters is the solution of the minimization of $J(\delta, W)$. Note that the number of moments M in the objective function has to be equal or greater than the number of parameters to be estimated, that is the dimension $D$ of $\delta$. If $M=D$ there is perfect identification and the solution of the minimization is the set of parameters $\delta_0$ such that $J(\delta_0, W)$=$0$. 
Note also that this condition holds necessarily only in the case in which $J(\delta, W)$ is continuous in $\delta$, which is not the case in computational model ($\delta$ is a discrete variable).
} 
 as we anticipated implying the observational equivalence. In such a case different set of structural parameters produce the same data distribution and it is a serious problem because economic models with different interpretations are completely indistinguishable; (ii) The under-identifiability  problem occurs when the objective function doesn't depend on certain structural parameter; (iii) If two or more parameters enter in the objective function only proportionally preventing the identifiability  of such parameters. Let call it partial-identifiability ; (iv) When the objective function has a unique minimum and all parameters enter the objective function separately but its curvature is small along certains dimensions. Canova call it weak identifiability . Let remark that, if the first three cases emerge intrinsically from the definition of the economic model, last case (iv) depends on the resolution of the data and may be improved.  \\

In the last decades a large literature tried to establish sufficient and necessary conditions upon the model for the identifiability. A simple but at the same time powerful theorem is formulated again by (\citet{canova2009back}). He stated that a sufficient condition to confirm identifiability hold when the objective function have a unique extremum at the true parameter vector and that it has to display "enough" curvature in all relevant dimension. We will see later when and how this condition extend to ABMs. \\



The "Canova" condition, despite the simplicity, hides many subtleties and it is not easy to establish if it is true or not. 
Many authors have developed theoretically this intriguing problem specializing on linear system and considering local and global identifiability. 

Linear or linearized problem, as linearized DSGE equations, permit in general to solve analytically the identifiability  problem and the only problem which may discourage it, is the high dimensionality of such a model. Fortunately more efficient numerical methods and very powerful large computer allows us to test identifiability in any linear or linearized DSGE models (\citet{le2013monte}). \\

Actually there is a considerable stir about extending this analysis to non linear systems in economics. (\citet{grewal1976identifiability}) demonstrated that local identifiability of the linearized form of a nonlinear model implies local identifiability of the non linear model. Nevertheless one cannot infer anything on the global identifiability and in addition to this not all non linear models are linearizable. 
On the other hand in non linear and complex model as well as the high dimensionality, we have to take into account complicated feedbacks between agents and between the micro and macro scale. This feedbacks in dynamical systems transcribe in non linear interactions and they formally prevent us to solve the dynamics analytically. The only way is to choose a numerical protocol which be as little time consuming as possible. \\

There exists several methods to face the identifiability  problem especially if the model is under identified; for instance a standard way is dropping out the non identified parameters or calibrating some of the parameters and estimating the others, conditional on the calibrated values. We could use directly real data in the model (stock market return for exemple) or use sensitivity analysis to select the parameters that make difference in the macro behavior of the model, fixing the other parameters to "reasonable" values. \\

In (\citet{canova2009back})  the authors suggests four strategies in addition to restrain the parameters set to a sub set of identifiable parameters: (i) explore the properties of the objective function; (ii) perform Monte Carlo samples, (iii) include in the objective function as many implications from the model as possible, and avoid the improper use of prior restriction, (iv) use estimation methods which are robust to identifiability  problems, or rethink the model specification and parametrization. Recently efficient numerical protocols are formalized to check global identifiability of DSGE in any dimension (\citet{le2013monte}).\\

A priori these solution are useless to be used with ABM first and foremost because it is not always possible to establish if the Canova's condition holds. Indeed there exists case in which the objective function doesn't have a minimum or the function displays a non differentiable shape. 


\vspace{1cm}

\section{Agent Based Modeling: a Recursive System Representation \label{sec:ABMrec}}


According to \citet{izquierdo} "...many computer models in the social simulations literature can be usefully represented as time-homogeneous Markov chain". Following this idea we study the problem of identifiability in ABM when the model can be seen as a Markov chain. 

Markov Chain are introduced by Andrey Markov (1907) as a random process usually characterized as memoryless. In particular the probability distribution of the next state depends only on the current state and not on the sequence of events that preceded it.

More formally a Markov sequence is a collection of random variables $\{X_j\}$ where $X$ is in general a real number and $j$$\in$$\mathbb{N}$, such that:
\begin{equation}
P(X_j = k | X_0 = p_0, X_1 = p_1, ..., X_{j-1} = p_{j-1}) = P(X_j = k | X_{j-1} = p_{j-1}) 
\end{equation}
In particular if a Markov sequence of random variables $X_n$ take the discrete values $a_1,..., a_N$, then:
\begin{equation}
P(x_n = a_{i_n} | x_{n-1} = a_{i_{n-1}}, ...,   x_{1} = a_{i_{1}}) = P(x_n = a_{i_n} | x_{n-1} = a_{i_{n-1}}) 
\end{equation}
and the sequence $x_n$ is called Markov chain (Papoulis 1984, pag 532).

The Markov chain approach helps in finding consistent estimators to consider in the objective functions. In particular an ergodic and aperiodic Markov chain admits maximum likelihood estimators (\citet{fabrettimarkov}).  On the other hand this is quite coherent with actual economic model which are mostly markovian and ergodic, except for some particular classes of economies which don't admit a unique statistical equilibrium (e.g. statistical multi-equilibria) . \\

Let us recall that a stochastic system is called ergodic if it tends in probability towards a limiting form that is independent of the initial conditions. In other words, with stationarity the moments are constant within the series, with ergodicity the moments are constant between the series. To test the ergodicity of the data produced by a model it is necessary to test whether the moments produced by different realizations tend towards the same value. \\

We restrain our analysis to this class of dynamical systems and consequently we propose ABM admitting the same properties, that is represented by a ergodic Markov chain. Therefore the system at time at time $j$ $\in$ $\mathbb{N}^*$  is given by the Markov chain of all micro states at time $j$, $X_j \equiv \{x_{ij}\}$, where $i \in$ $\mathbb{N}^*$ represent each agent. \\

Recently \citet{leombruni2005economists, richiardi2006common, gallegati2011agent} have shown that an AB model can be seen as a recursive system. This is not surprising because an ABM is an adaptative automata that at each time step updates its local rules.  The evolution state variable $x_{i,j}$ is specified by the finite difference equations:
\begin{equation}
x_{i,j+1} = f_{i}(x_{i,j},x_{-i,j};\theta_i; (1-\eta)\xi)
\label{eq:transitionEQ1}
\end{equation}
where $f_i$ is a function taking value in $\mathbb{R}^{k}$, $\theta_i$ are the vector of parameter for each agent. Here we assume that the behavioral rules \footnote{Let us remark that we use "behavioral rules" in a loose sense that encompasses the actual intentional behavior of individuals as well as environmental factors such as technology. The agents are in principles heterogeneous and ables to adapt their-self to the output.} may be individual-specific both in the functional form of the phase line $f_i(\cdot)$ and in the vector parameter $\theta_i$  and may be also based on the state $x_{-i}$ of all individuals other than $i$. In the following we will consider discrete time dynamics because they are more consistent with numerical simulation; nevertheless no conceptual difficulties prevent us to recast all the formalism in continuous time or taking the continuous limit of the model. \\

If we define $\Xi_{j} \equiv (1-\eta)\xi_{i,j} $ the vector of stochastic elements \textit{i.i.d.} and $\Theta$ the parameter vector $\theta_i$, $i \in$ $\mathbb{N}^*$  we can recast (\ref{eq:transitionEQ1}) in the more convenient form and compact form:
\begin{equation}
x_{i,j+1} = f_{i}(X_j;\Theta; \Xi)
\label{eq:transitionEQ2}
\end{equation}

The stochastic term $\Xi_{j}$ is necessary to characterize the field $\textbf{X}_j$ as a markovian process; it can be seen for instance as a white noise, that is real valued i.i.d. random variables. In order to keep the model as general as possible note that $\eta$ change the amplitude of the noise and for $\eta=1$ recover a purely deterministic data generating process.  \\

Now let consider the dynamical system but for the entire state $X_j$ = $(x_{i,j})$, the tensor of dimension ($N_{max} \times 1$) where $N_{max}$ is the maximum number of agents (possible infinite). This Markov chain with a multidimensional index is called Markov Random Field (MRF). For each time $j$ the MRF obeys to:
\begin{equation}
X_{j+1} =  F(X_j;  \Theta; \Xi)
\label{eq:transitionEQ3}
\end{equation}
Note that this formalism and in particular eq. (\ref{eq:transitionEQ3}) is the same as the one adopted to analyze DSGE models where the vector $\Xi_j$ is the vector of innovations including the random shocks, which are the engine of the dynamics. However while in DGSE the transition function $F(\cdot)$ has an explicit analytical formulation, in ABM it remains implicitly defined by the micro-transition equations (\ref{eq:transitionEQ2}). In general let call eq. (\ref{eq:transitionEQ3}) the \textit{Data Generating Process} (DGP) equation.\\

As in many practical cases we are interested in a specific aggregate observable $Y_j$, such as $GDP$, or unemployment rate, $etc$; we therefore measure this observable projecting the state $X_j$ to $Y_j$ and express:
\begin{equation}
Y_j= G(X_j; \Theta; \Xi) 
\label{eq:measureEQ}
\end{equation}
Remark that in practice in numerical simulation random terms are not really random but pseudo-random, because classical computers are deterministic. We can therefore describe  measurement equation as $Y_{j}= G_j(Z_0; \Theta)$ where $Z_0 = \{X_0,s\}$ where $s$ is the random seed.

Equation (\ref{eq:measureEQ}) is also present in DSGE modeling as the measurement equations and with the transition equations form together the state space representation of the system. \\

As before in the DSGE case the functional form of $G(\cdot)$ may be analytically known, but it is not the case in ABM. A simpler and less abstract way to see at this problem is by iteration on each terms $x_{j,m}$ of $X_j$:
\begin{equation}
Y_0 = G(x_{1,0},...,x_{n,0})  \nonumber
\end{equation}
\begin{equation}
Y_1 = G(x_{1,1},...,x_{n,1}) =G(F_1(x_{1,0}x_{-1,0};;\theta_1),...,F_n(x_{n,0}x_{-n,0};;\theta_n)) \equiv G_1(x_{1,0},...,x_{n,0};\theta_1,...\theta_n)  \nonumber
\end{equation}
\begin{equation}
... \nonumber
\end{equation}
\begin{equation}
Y_n = G_n(x_{1,0},...,x_{n,0};\theta_1,...\theta_n)  \nonumber
\end{equation}
We call eq. ($\ref{eq:measureEQ}$) the \textit{Input Output Transformation} (IOT) equation because the laws of motion uniquely relates the value of $Y$ at any time $j$ to the initial conditions of the systems and to the value of the parameters.\footnote{Equation (\ref{eq:measureEQ}) can be solved alternatively studying for all $(n, j) \in {\mathbb N}^2$, the collection 
$W_j^n = (Y_{k})_{k = nj}$ or equivalently recursively. This collection represents the aggregate statistics $Y$ at `time' $k=n j$. For any given $n$, the collection $S^n = (W_j^n)_{j \in \mathbb Z}$ thus represents the entire history of the statistics observed through a stroboscope of `period' $n$. These can be deduced from the original evolution equations (\ref{eq:measureEQ}) of the statistics, which also coincide with the evolution equations of $S^1$.  The evolution equations for the statistics $Y_j$ after a period of $n$ is given by $S^n$ which represent the link between $W_{j+1}^n$ to $W_j^n$ for all $j$}.\\

\subsection{Ergodicity versus stationarity}

Let us recall that the transition matrix in a time homogeneous Markov process is defined as:
\begin{equation}
P(x,y) = Pr(X_{j+1} = y | X_j = x)
\end{equation}
The distribution after $N$ step is defined in the following way:
\begin{equation}
P^N(x,y)=\begin{cases} P(x,y), & \mbox{if } t =1 \\ \sum_{z\in\Omega}P(x,z)P^{N-1}(z,y), & \mbox{if }t > 1
\end{cases}
\end{equation}
where $\Omega$ is the state space. As we recalled in the first section, the ergodic Markov chains have a unique stationary (i.e. limiting) distribution. In particular we say a distribution $\pi$ is a stationary distribution if it is invariant with respect to the transition matrix, i.e.,
\begin{equation}
 \mbox{for all y} \in \Omega, \ \pi(y) = \sum_{x\in \Omega} \pi(x) P(x,y)
\end{equation}
In particular let us recall the useful theorem: \textit{For a finit ergodic\footnote{Note that the property to be irreducible or aperiodicity in Markov Chain are equivalent to ergodicity. In particular, a Markov chain is irreducible if \textit{for all x,y $\in \Omega$, there exists $t=t(x,y)$ such that $P^j(x,y)>0$}.} Markov chain, there exists a unique stationary distribution $\pi$ such that}:
\begin{equation}
 \mbox{for all x, y} \in \Omega, \ \lim_{j \to \infty}P^j(x,y) = \pi(y)
\end{equation}

The notion of equilibrium here is very different from what we call equilibrium in DSGE model. In DSGE models, according to the rational expectation paradigm, equilibrium is defined as a consistency condition in the behavioral equations, \textit{i.e.} agents (whether representative or not) must act consistently with their expectations and the action of all the agents must be mutually consistent. The condition logically operates at an individual level before aggregation and the system is therefore always in equilibrium, even during a phase of adjustment to a shock. \\

ABM, on the other hand, are characterized by (more or less sophisticated) adaptive expectations, according to which consistency might or might not arise, depending on the evolutionary forces that shape the system. An equilibrium can therefore be defined only at the aggregate level and only in statistical terms, after the macro outcomes have been observed. \\

In the special case in which the dynamical system is ergodic, the model will always ends up with the same statistical equilibria for any choices of the initial conditions and only depends on the parameter set. On the contrary if the system is not ergodic but stationary, for the same values of parameters, the system can end up with different equilibria and they depend on the initial conditions. 

\section{Estimation in ABM and simulation-based econometrics}

The aim of this section is to introduce briefly the reader to the different methods dealing with empirical validation of ABM in economics. In general what we do is not different from estimation of other economical non linear or linear model: we aims at inferring the true value of the parameters from the real data, by maximizing the fitness of the model with the observed data in a carefully chosen distance metric, such that the estimator has well known (at least asymptotically) properties. \\ 

On the other hand contrarily to analytical model, in ABM the high numbers of parameters and non-linearities, together with the fact that ABM cannot be solved analytically, leads to burdensome computational methods. \\

As for analytical models where the likelihood function is too complicated to be derived, estimation in AB models can be performed by means of simulations based techniques. The basic idea with simulation-based econometrics is to replace the evaluation of analytical expression about theoretical quantities with their numerical counterparts computed on the simulated data. The simulated theoretical quantities, which are functions of the parameters to be estimated can then be compared with those computed on the real data as in any estimation procedure. In asymptotic limit the observed quantities tend to the theoretical quantities, at the true values of the parameters. Because the simulated quantities also tend to the theoretical quantities, the observed quantities converse to the simulated quantities. Finding the values of the parameters that minimize the distance between the simulated and the observed quantities should therefore provide consistent estimates of the parameters. This procedure is generally known as indirect estimation.  \\

An important feature of estimation is consistency. We can list three kind of requirements in this respect: (i) consistency in size, when the estimated converge to their true value as the observed population of agents grows bigger; (ii) consistency in time (longitudinal consistency) means that the estimates converge to their true value as the length of the observation period increases; and finally (iii) consistency in replications as the convergence is faster as more occurrences of the same stochastic process are observed. \\

It is quite clear that an important assumption to obtain convergence is that the statistics used identify the parameters of interest, that is there is one to one relationship between the theoretical values of the statistics and the value of the parameters. 

\section{Two numerical tests of identifiability for ABM: problems and perspectives}

The aim of this section is to review two different ways for test identifiability in a specific class of ABM presented in the last section, introduced in other contexts by \citet{canova2009back} and \citet{richiardi2006common}. We present first (i) an extension of the Canova condition and Ricchiardi results employing a simulated method distance protocol and a bayesian protocol which generalize the Canova condition to a posterior distribution; (ii) an indirect inference method.

\subsection{1a. Simulated minimum distance}

As we've discussed in the third section the relationship between the behavior of ABM and the structural parameters remains hidden in the recursive structure of the model. Then only inductive evidence can be obtained by analyzing the simulated data. This implies that it is not possible to write analytically the objective function and that the properties of the statistics used in the objective functions are also not analytically known. Simulation based econometric methods overcome this problem by simulating the model and using the artificial data to estimate $\Theta$ as anticipated in previous section.\\

Let us recall briefly the minimum distance works by comparing theoretical constructs (e.g. longitudinal moment) computed over $Y_j$, $\mu(Y_j(\Theta))$, which depends on the structural parameter $\Theta$, with their observed counterparts $\mu^R$, computed on $Y^R$. As usual we can find a $\Theta$ in order to minimize the distance between observed data and theoretical ones. More formally,  define $\hat \mu_n$ a set of estimator and $\mu(\Theta)$ the mapping between the model and the estimators; the objective function transcribes:
\begin{equation}
\hat Q_{(\Theta)} = - (\hat \mu_n - \mu(\Theta))'  \hat W_n(\hat \mu_n - \mu(\Theta))
\label{eq:Q}
\end{equation}
where $\hat W_n$ is a positive semi-definitive matrix. Minimizing the distance between real data and theoretical one means minimizing the latter (\ref{eq:Q}), that is:
\begin{equation}
\hat \Theta_n = \text{argmax}_\Theta \hat Q_n (\Theta).
\label{eq:th}
\end{equation}

In simulated minimum distance we must extend the minimum distance estimation by replacing the theoretical properties of the model $\mu(\Theta)$ with their simulated counterpart $\tilde \mu(\Theta, \epsilon_s)$ where  $\epsilon_s$ denotes the random term in the simulated data. According to our technical computational possibilities we can simulate all configuration of the objective function for all $\Theta$. This allows us solve numerically, if it is possible, the maximization problem.\\
\begin{figure}[h!]
\centering
\includegraphics[width=1\columnwidth]{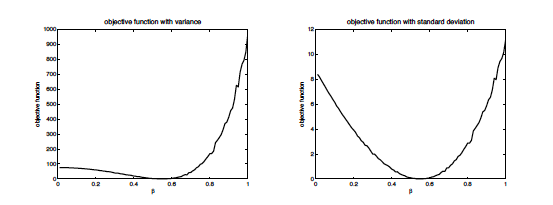}
\caption{Ricchiardi and Grazzini (2014), Objective function, constructed over different momenta. Left: Variance. Right: standard deviation. The theoretical moments are computed, for each value of the only parameter, averaging over 100 runs each lasting for 400 trading day after the absorbing equilibrium is reached. The Model used here is an AB stock market proposed in Cliff and Bruten, 1997 to reproduce the experimental results obtained by Smith (1962).}
\label{fig:fig1}
\end{figure}

\begin{figure}[h!]
\centering
\includegraphics[width=1\columnwidth]{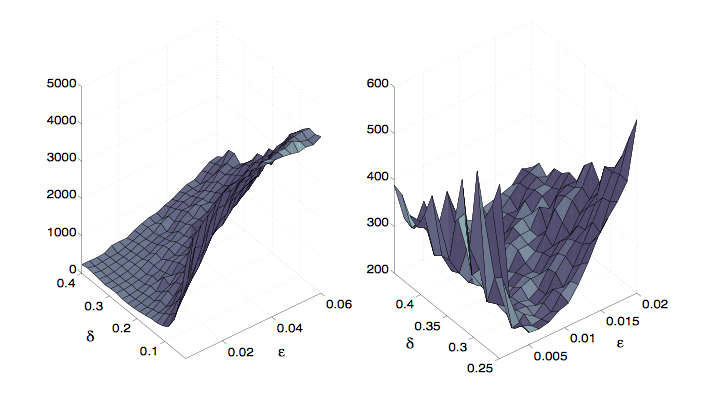}
\caption{Winker, Gilli, Jeleskovic (2006) Objective function for the mean for a large number of Monte-Carlo replications (100) with regards to a subset of parameter ($\epsilon$ and $\delta$). The model used here is the Kirman's model. Based on the findings in the left plot, a second simulation has been run on a smaller subset of the parameter space comprising those values which appeared most favorable in the first run.}
\label{fig:fig1}
\end{figure}
In other terms, if consistency conditions to estimate an ABM by SMD hold, we can in principle (i) entirely reproduce (\ref{eq:Q}); (ii) the ergodicity assure that final statistical distribution doesn't depend on the initial condition and that and (iii) the markovianity together with the ergodicity assure that the limiting distribution is unique. In principle we can check if the (\ref{eq:Q}) admits a global minimum which implies that the model is identified respect to $\Theta$.  In Fig. \ref{fig:fig1} we can see that there exists a global minimum for the model proposed by Ricchiardi and Grazing (2014). The Model used here is an AB stock market proposed in Cliff and Bruten, 1997 to reproduce the experimental results obtained by Smith (1962). Let remark that the curvature of the objective function is smaller for the mean (linear momentum) that for the standard deviation (quadratic momentum) displaying a simple case of partial-identifiability (\citet{canova2009back}). \\
Less computer demanding is checking a local identifiability. This can be done by a perturbation method looking for a local minimum in a volume $d\Omega$ of dimensions $(\theta_1,\theta_2,....\theta_n)$. An exemple is shown in the Winker, Gilli, Jeleskovic (2006) paper where the objective function of the mean regards to a subset of parameters displays a minimum in a smaller subset of those 2 parameters. 


\subsection{1b. Bayesian estimation}

Another way to estimate and test identifiability in ABM is through Bayesian estimation.  The fundamental equation for Bayesian methods is a simple application of the Bayes theorem:
\begin{equation}
p(\theta | Y^R) \propto \mathcal{L}(\theta; Y^R) p(\theta)
\end{equation}
where $p(\theta)$ is the prior distribution of the parameter, $\mathcal{L}(\theta; Y^R)  \equiv p(Y^R | \theta)$ is the likelihood of observing the data $Y^R  \equiv \{y^R_{j} \}$, given the value of the parameters, and $p(\theta | Y^R)$ is the posterior distribution, that is the updated distribution once the information coming from the observed data is properly considered.\\

Sampling the posterior distribution $p(\Theta | Y^R)$ involves two computationally intensive steps: (i) for given value of $\theta$, obtaining the likelihood $\mathcal{L}$ and (ii) iterating over different values of $\theta \in \Theta$. The latter is also needed for SMD where each value of $\theta$ is associated with a score which measures the distance between the theoretical and observed quantities.\\
In simple model the exploration of the parameter space can be accomplished by "brute force" ride iteration: the parameters space is sampled at regular (small) intervals. However, such a systematic search is highly inefficient, as it involves evaluating the density of the posterior distribution at many points where it is practically zero, while more likely values of $\theta$, where a finer search might be valuable, are sampled with the same probability.  More efficient Markov Chain Monte Carlo or genetic algorithm can then be used (\citet{fabrettimarkov}).\\

Assuming ergodicity and (weak) stationarity, the probability of observing the whole (unordered) series of data $Y^R$ is simply:
\begin{equation}
\mathcal{L}(\theta, Y^R) \propto \prod_{j=1}^T f(y_j^R | \theta)
\end{equation}
where we assume "to be blind" with respect to all the other data points $\{y_{-j}^R \}$ where we evaluate $y_{j}^R$. 


In order to compute the likelihood all we need is an estimate for the distribution $\tilde f(\theta)$. We then evaluated the estimated distribution  $\tilde f(\theta)$ at each observed $y_j^R$ and compute $\prod \tilde f(y_j^R|\theta)$. The estimation of the density, for any value of $\theta$, is done by simulation, that is, by repeatedly running the model with those parameters and observing the outcomes. Let remark that if the outcomes $Y$ were discrete, we would only have to count the frequency of occurence of each observed value $y_j^R$. 


Here in order to test identifiability, we can check numerically if the posterior distribution admits a global minimum or local minima in an analogous way of simulated objective function in SMD . Note that compute posterior distribution is computational intensive even if the numerical code is parallelized and in conclusion the SMD could be easier and faster to implement.



\subsection{2. Indirect inference}

Indirect inference is largely employed in linear and non linear model for estimation. The basic idea of the indirect inference for estimation is to use the coefficients of an auxiliary model, estimated both on the real and on the simulated data, to describe the data, that is as summary statistics on the original model. The method prescribes the following steps: (i) simulate the model for given values $\theta_0$ of the parameters and obtain artificial data; (ii) estimate the parameters of a auxiliary model; (iii) change the structural parameters of the original model until the distance between the estimates of the auxiliary model using real and artificial data is minimized. \\
Le, Minford and Wickens (2013) (\citet{le2013monte}) have recently proposed a Monte Carlo procedure, based on an indirect inference method, for checking identification in DSGE models. The test consisted in search for different set of structural parameters that could potentially also generate the auxiliary model's parameters. In the affirmative case the model was not identified. The fundamental assumption for this procedure was the linearity of the model and consequently that a local identification implied a global identification. \\
For a non linear model local identification doesn't imply global identification but it is still possible to check identification for a subset of parameters. 

\section{A mean field approach?}

As a consequence of the dissatisfaction with the RA approach on one hand and on the other hand the complexity arising from computational economics like ABM, a few analytical frameworks have been developed. One of the most promising methods has been introduced by Duncan Foley and Masanao Aoki  ( \citet{aoki1998new, aoki1994new, aoki2001modeling}) who borrowed from statistical mechanics the concept of mean-field interaction and imported it into economics. In the mean-field interaction approach, agents are classified into clusters or sub-systems according to their state with respect to one particular feature (the microstate). This clustering determines the characteristics and the evolution of the aggregate (the macrostate). The focus is not more on the single agent, but on the number or fraction of agents occupying a certain state of a state-space at a certain time. 
The analytical solution to ABM is the result of the functional-inferential method which identified the most probable path of the system dynamics. The method considers the heterogeneity, representing a large number of agents, and the interaction among them, which originates fluctuations of the macroeconomic variables about a deterministic trend. Individual direct interaction is replaced by indirect mean-field interaction between sub-systems, expressed in terms of the transition rates of the master equations. 
From master equations we can compute analytically the equilibrium distributions, that is the statistical stationary equilibrium. For an interesting application on financial markets see for exemple \citet{gatti2012reconstructing}). \\

An other similar method is to obtain master equations directly from the Langevin-like equation (\ref{eq:transitionEQ3}). This discrete time dynamical system may admits a continuous limit if we consider a time interval $\Delta t$ $<<$ 1,$ t_j = j \Delta t$ and the Markov chain is homogeneous in time. Then letting $\Delta t$ to zero, the formal continuous limit is:
\begin{equation}
\partial_t X_t = \tilde F(X;\Theta;\Xi) 
\label{eq:ME}
\end{equation}
where $\tilde F(\cdot)$ takes into account the first order in O($\Delta t$). This equation is known as Langevin equation for a multi-component process $X$ = $(x_1,x_2, ... , x_N)$ describing the time evolution of macroscopic states under the action of microscopic agents dynamics. It is well known from statistical physics that the Kramers-Moyal expansion of the master equation (\ref{eq:ME}) includes up to second-order terms define a probability distribution obeying a Fokker Plank (FP) equation. The latter has the following form:
\begin{equation}
\partial_t P = - \sum_i \partial_{x_i} \left( \Omega(X, \Theta) P \right) + D \sum_{ij}\partial_{x_i,x_j}\left( \sum_k \Lambda_{k i} \Lambda_{k j} P \right)
\end{equation}
where $\Omega$ and $\Lambda$ depends on the functional form of $\tilde F(\cdot)$. Let us recall that the knowledge of the FP equation permits to compute the stationary probability distribution analytically. \\

Mean field approximation can be also obtained through coarse-graining. In other words, let consider the system in different levels of detailing and suppose that one zoom out from the microscopic description. By doing so, one averages over the possible microscopic descriptions which are compatible with a higher level description. The microscopic correlations are gradually suppressed but the emergent statistics of homogeneous agent take into account initial heterogeneity.   The use of ABM is particularly useful for this coarse graining procedure, as its discrete structure gives a clear visualization of the process: instead of being able to resolve a single agent path, one is only able to see bigger blocks, thus not taking into account the full information about the microscopic structure. This erasure of information will lead to an effective state and effective dynamics, which will be possibly simpler than the microscopic one.\\

Why mean-field method, coarse-graining or Langevin approach are useful for testing identification? Master equations are a closed form equation from which we can derive analytically the objective function at least at first order around the stationary probability $P_{st}$. \\

\citet{gualdi2015tipping} show recently that in some cases is possible to recognize in the rules of ABM those conditions getting stationary points (that is equivalent in ergodic system to a statistical equilibrium). In their paper the authors explores the possible type of phenomena that a simple macroeconomic AB model can reproduce. They found analytically that their model admits a stationary state $X_{st}$ and for analytical purpose they focus in limit of small oscillation around the stationary point. 
Remarkably they compute the associate discrete time evolution of the probability distribution, in other words the effective master equation of ABM evolution. Then they compute the stationary probability $P_{st}$.  Knowing the stationary probability analytically we are able to reconstruct analytically the objective function for any momentum and verify if it admits minima or not. \\ 
Again this is not in general feasible but when it is possible displays an alternative answer to simulated econometrics.

\section{Discussion}

We've shown that markovian and ergodic ABMs can be recast in a simple and formal way in order to map the reduced form of any DSGE model. For both model the estimation and consequently the identifiability test of structural parameters, is an hard problem. This is a straightforward consequence of non linear terms and in some cases of the high dimensionality of the system. \\

In general we cannot test analytically if a DSGE or an ABM are identified except in some very specific cases: for instance if the moment conditions are equal to or greater than the parameters to be estimated, they are linearly independent and monotonically increasing or monotonically decreasing in all the parameters, than the objective function has a unique maximum. Nevertheless to be useful this very intuitive result it is necessary to know analytically the objective function and it is quite never the case. \\

The results obtained for linearized DSGE are not useful for ABM because (i) non linear orders are fundamental to observe a non trivial dynamics, (ii) due to the complex behavior of ABM linearizing equations for each agents is not the same that linearizing dynamical equation for the aggregate, (iii) however this is not feasible due to the hight dimensionality of agents state space. \\ 

In case of markovian and ergodic dynamical system the existence of a unique statistical equilibrium implies the existence of a stationary probability. This seems the key to threat the problem of identifiability. In fact, (i) the ergodicity means that the final stationary probability doesn't depend on the initial condition but only on the parameter set. This simplify enormously a numerical reconstruction of the objective function and its minimization. (ii) It is possible to write the master equations about the stationary probability (and assuming the small oscillation). This leads us to a closed form equation and a analytical expression of the stationary probability. It is than straightforward computing the analytical form of the objective function for each momentum.  \\

Identifiability is a fundamental and not only technical problem in economic dynamic model. ABMs and non linear DSGE researchers shares the same hard mission to find numerical protocol more optimized and efficient. We believe, however, that analytical test are also possible in some special, but highly representative range of solutions, for instance developing further and deeper a mesoscopic statistical analysis of ABMs.\footnote{I would thank Matteo Ricchiardi for most useful discussions especially during the MAGECO workshop on April 2015 in Paris.}

 \end{large}


\newpage

\appendix


\bibliographystyle{plainnat}

\bibliography{abm}

\end{document}